\documentclass[a4paper,12pt]{article}
\usepackage{amssymb}
\usepackage{latexsym}
\usepackage[dvips]{graphicx}
\newcommand{\be}{\begin{equation}}
\newcommand{\ee}{\end{equation}}
\newcommand{\bea}{\begin{eqnarray}}
\newcommand{\nn}{\nonumber}
\newcommand{\eea}{\end{eqnarray}}
\begin{document}

\begin{titlepage}
\begin{flushright}
hep-th/0109108\\ UA/NPPS-11-2001
\end{flushright}
\begin{centering}
\vspace{.8in}
{\large {\bf Semiclassical Corrections\\ to the Bekenstein-Hawking entropy
\\of the BTZ Black Hole via Self-Gravitation}}
\\

\vspace{.5in}
{\bf Elias C. Vagenas\footnote{hvagenas@cc.uoa.gr}} \\

\vspace{0.3in}
University of Athens\\ Physics Department\\
Nuclear and Particle Physics Section\\
Panepistimioupolis, Ilisia 157 71\\ Athens, Greece\\
\end{centering}

\vspace{1in}
\begin{abstract}
\noindent 
Hawking radiation is viewed as a tunnelling process. In
this way the effect of self-gravitation gives rise to semiclassical
corrections to the entropy of the ($2+1$) BTZ black hole. The
modified entropy, due to specific modelling of the 
self-gravitation effect, of the ($2+1$) BTZ black hole is evaluated.
To first order in $\omega$ which is a shell of energy radiated 
outwards the event horizon of the BTZ black hole, modified entropy is 
proportional to the horizon. 
In this semiclassical analysis, corrections to the 
Bekenstein-Hawking formula $S_{BH}=\mathcal{A}_{H} /
4l_{P}^{2}$ are found to be negative and the proportionality factor connecting
the modified entropy, $S_{bh}$, of the ($2+1$) BTZ black hole 
to the Bekenstein-Hawking entropy, $S_{BH}$, is 
evaluated to first order in $\omega$.
\\
\\
\\
\\
\\
\\
\\
\end{abstract}

\end{titlepage}
\newpage

\baselineskip=18pt
\par\noindent
In this work we apply the  Keski-Vakkuri, Kraus and Wilczek
 \cite{KKW1,KKW2,KKW3,KKW4} methodology to the ($2+1$) BTZ
black hole. The idea of KKW method is to view the BTZ black hole
background as dynamical by treating the Hawking radiation
\cite{hawking} as a tunnelling process
\cite{correction1,correction2,correction3,correction4,correction5}. 
The energy conservation is the key to this description. 
The total (ADM) mass is kept fixed while the mass of the 
BTZ black hole decreases due
to the emitted radiation. The effect of self gravitation gives
rise to additional terms in the Bekenstein-Hawking area  formula
concerning the BTZ black hole entropy.
\par \noindent
The black hole solutions of Ba$\tilde{n}$ados, Teitelboim and
Zanelli \cite{banados1,banados2} in ($2+1$) spacetime dimensions
are derived from a three dimensional theory of gravity: \be S=\int
dx^{3} \sqrt{-g}\,({}^{{\small(3)}} R+2\Lambda) \ee with a
negative cosmological constant ($\Lambda>0$). 
\par\noindent
The corresponding
line element is: \be ds^2 =-\left(-M+\Lambda r^2 +\frac{J^2}{4
r^2} \right)dt^2 +\frac{dr^2}{\left(-M+\Lambda r^2 +
{\displaystyle \frac{J^2}{4r^2}} \right)}
+r^2\left(d\theta -\frac{J}{2r^2}dt\right)^2
\label{metric}\ee with $M$ the ADM mass, $J$ the angular momentum
(spin)
 of the BTZ black hole and $-\infty<t<+\infty$, $0\leq r<+\infty$.
\par \noindent
For the positive mass black hole spectrum with spin ($J\neq 0$),
the line element (\ref{metric}) has two horizons: \be
r^{2}_{\pm}=\frac{M\pm\sqrt{M^2 - \Lambda J^2} }{2\Lambda}
\label{horizon1} \ee with $r_{+}$, $r_{-}$ the outer and inner
horizon respectively. 
\par\noindent
The area $\mathcal{A}_H$ and Hawking
temperature $T_H$ of the event (outer) horizon are
\cite{carlip1,carlip2}: \bea \mathcal{A}_H
&=&2\pi\left(\frac{M+\sqrt{M^2 -\Lambda J^2
}}{2\Lambda}\right)^{1/2}\nn\\ &=&2\pi r_{+} \\ T_H &=&
\frac{\sqrt{2\Lambda}}{2\pi}\frac{\sqrt{M^2-\Lambda J^2}}
{(M+\sqrt{M^2 -\Lambda J^2})^{1/2}}\nn\\
&=&\frac{\Lambda}{2\pi}\left(\frac{r_{+}^{2}-r_{-}^{2}}{r_{+}}\right)\, .
\label{temp1}\eea 
The entropy of the spinning BTZ black hole is:
\be S_{bh}=4\pi r_{+} \label{entropy1}\ee and if we reinstate the
Planck units (since in BTZ units $8\hbar G =1 $) we get: 
\be
S_{bh}=\frac{1}{4\hbar G} \mathcal{A}_H =S_{BH}
\ee 
which is the
well-known Bekenstein-Hawking area formula ($S_{BH}$) for the entropy
\cite{bekenstein1,bekenstein2,hawking,hawking1} (or
\cite{strominger3} where the formula is proven by counting excited
states).
\par \noindent
In order to apply the idea of Keski-Vakkuri, Kraus and Wilczek
 to the positive mass BTZ black
hole spectrum with fixed $J$ we have to make a coordinate
transformation. We choose the Painlev$\acute{e}$ coordinates
\cite{painleve} (utilized recently in \cite{wilczek} for black
hole backgrounds) which are non-singular on the outer horizon
($r_{+}$). We are thus able to deal with phenomena for which the
main contribution comes from the outer horizon.
\par\noindent
We introduce the time coordinate $\tau_P$ and the angular
coordinate $\theta_{P}$ by imposing the ans\"{a}tz:
\bea
 \sqrt{g(r)}\,dt &=& \sqrt{g(r)}\,d\tau_{P}- \frac{\sqrt{1-g(r)}}{\sqrt{g(r)}}dr \\
\sqrt{g(r)}\,d \theta  &=& \sqrt{g(r)}\,d \theta_{P}-\left( \frac{J}{2 r^2}
\right) \frac{\sqrt{1-g(r)}}{\sqrt{g(r)}} dr
\eea
where
\be
g(r)=-M+\Lambda r^2 +\frac{J^2}{4r^2} \, .
\ee
The line element (\ref{metric}) is now written as: \bea ds^2 & =
& -g(r) d\tau^{2}_{P}+2\sqrt{1-g(r)}d\tau_{P}dr +dr^2
+\frac{J^2}{4 r^2}d\tau^{2}_{P}-Jd\tau_{P}d\theta_{P} +r^2
d\theta^{2}_{P}\\ &=&-\left(-M+\Lambda r^2\right)d\tau_P^2
 + 2 \sqrt{1+M-\Lambda r^2-\frac{J^2}{4 r^2}}\,d\tau_P dr + dr^2\nn\\
&&+\,r^2 d\theta^{2}_{P} - Jd\tau_{P} d\theta_{P}\, .
\label{metric3}
 \eea
It is obvious from the
above expression that there is no singularity at the points  $r_+$
and $r_-$. The radial ($\theta=constant$) null ($ds^2 =0$) geodesics followed by a
massless particle are:
\be
\dot{r}\equiv\frac{dr}{d\tau_P}=\pm \sqrt{1-\frac{J^2}{4 r^2}}-
\sqrt{1+M-\Lambda r^2-\frac{J^2}{4 r^2}} \ee where the signs $+$
and $-$ correspond to the outgoing and ingoing geodesics,
respectively, under the assumption that $\tau_P$ increases towards
future.\par \noindent
 We next fix the total ADM mass and let the
mass $M$ of the BTZ black hole vary. If a shell of energy (mass)
$\omega$ is radiated outwards the outer horizon then the BTZ black
hole mass will be reduced to $M-\omega$ and the shell of energy
will travel on the modified geodesics:
\be \dot{r}=+\sqrt{1-\frac{J^2}{4 r^2}}-
\sqrt{1+(M-\omega)-\Lambda r^2-\frac{J^2}{4 r^2}}
\label{geodesic}\ee produced by the modified line element:
\bea
ds^2 &=& -\Bigg(-(M-\omega)+\Lambda r^2\Bigg)d\tau_P^2
 + 2 \sqrt{1+(M-\omega)-\Lambda r^2-\frac{J^2}{4 r^2}}\,d\tau_P dr + dr^2\nn\\
&&+\,r^2 d\theta^{2}_{P} - Jd\tau_{P} d\theta_{P}\, .
 \label{metric4}
\eea
\par\noindent
It is known that the emission rate from a radiating source can be
expressed in terms of the imaginary part of the action for an
outgoing positive-energy particle as: \be \Gamma=e^{-2ImS}
\label{gamma}\ee but also in terms of the temperature and the
entropy of the radiating source which in our case will be a BTZ
black hole: \be \Gamma=e^{-\beta \omega}=e^{+\Delta
S_{\hspace{0.05cm}bh}} \label{rate1} \ee where $\beta$ is the
inverse temperature of the BTZ black hole and $\Delta
S_{\hspace{0.05cm}bh}$ is the change in the entropy of the BTZ
black hole before and after the emission of the shell of energy
$\omega$ (outgoing massless particle). It is clear that if we
evaluate the action then we will know the temperature and/or the
change in the entropy of the BTZ black hole. 
\par\noindent
We therefore evaluate
the imaginary part of the action for an outgoing positive-energy
particle which crosses the event horizon outwards from:
 \be
r_{in}^2=r_{+}^2(M, \Lambda, J)=\frac{M+\sqrt{M^2 - \Lambda J^2}
}{2\Lambda}
\label{horizonin}
\ee 
to 
\be 
r_{out}^2=r_{+}^2(M-\omega
,\Lambda, J)=\frac{(M-\omega)+\sqrt{(M-\omega)^2 - \Lambda J^2}
}{2\Lambda}\, .
\label{horizonout} 
\ee
The imaginary part of the action is: 
\be
ImS=Im\int_{r_{in}}^{r_{out}}p_{r}dr=Im\int^{r_{out}}_{r_{in}}
\int_{0}^{p_{r}}dp'_{r}dr\, . 
\ee 
We make the transition from the
momentum variable to the energy variable using Hamilton's equation
$\dot{r}=\frac{dH}{dp_{r}}$  and equation (\ref{geodesic}). The
result is:
\be
ImS=Im\int^{r_{out}}_{r_{in}}\int^{\omega}_{0}\frac{(-d\omega')dr}
{\sqrt{1-{\displaystyle\frac{J^2}{4 r^2}}}-\sqrt{1+(M-\omega')-
\Lambda r^2 -{\displaystyle \frac{J^2}{4 r^2}}}}
\ee where
the minus sign is due to the Hamiltonian being equal to the
modified mass $H=M-\omega$. This is not disturbing since
$r_{in}>r_{out}$. After some calculations (involving contour
integration into the lower half of $\omega'$ plane) we get:
\be
ImS=2\pi\Bigg\{\Big[r_{in}\sqrt{1-\frac{J^2}{4 r^{2}_{in}}}+
\frac{J}{2} \arcsin(\frac{J}{2 r_{in}})\Big]-\Big[r_{out}\sqrt{1-
\frac{J^2}{4 r^{2}_{out}}}+\frac{J}{2} \arcsin(\frac{J}{2
r_{out}})\Big]\Bigg\}. \ee Apparently the emission rate depends
not only on the mass $M$ and angular momentum (spin) $J$ of the
BTZ black hole but also on the energy $\omega$ of the emitted
massless particle:
 \bea
\Gamma(\omega,M, \Lambda, J)=e^{-2ImS}\hspace{11cm}&&\nn\\
= exp\left[ 4\pi
\Bigg\{\Big[r_{out}\sqrt{1-\frac{J^2}{4 r^{2}_{out}}}+\frac{J}{2}
 \arcsin(\frac{J}{2 r_{out}})\Big]-
\Big[r_{in}\sqrt{1-\frac{J^2}{4 r^{2}_{in}}}
 +\frac{J}{2} \arcsin(\frac{J}{2 r_{in}})\Big]\Bigg\}\right].
&&\label{temp2}
\eea
Comparing
(\ref{temp2}) and (\ref{rate1}) in which:
\be
\Delta S_{bh}=S_{bh}(\omega,M, \Lambda, J)-S_{bh}(M, \Lambda, J)
\label{entropy}
\ee
we deduce that the modified
entropy due to the specific modelling of the 
self-gravitation effect of the BTZ black hole is:
\be
S_{bh}(\omega,M, \Lambda, J)=4\pi r_{out}\sqrt{1-\frac{J^2}{4
r^{2}_{out}}}+2\pi J\arcsin\left(\frac{J}{2 r_{out}}\right)+S_{0}
\label{mentropy1} 
\ee 
where $S_{0}$ is an arbitrary constant which 
will be evaluated. 
\par\noindent
The second term in equation (\ref{entropy}) should be
the Bekenstein-Hawking area formula:
\be
S_{bh}(M, \Lambda, J)=S_{BH}
\ee 
and comparing with equation 
(\ref{temp2}) we obtain:
\be
S_{BH}=4\pi r_{in}\sqrt{1-\frac{J^2}{4 r^{2}_{in}}}
 +2\pi J \arcsin(\frac{J}{2 r_{in}})+S_0 \, .
\ee
\par\noindent
Hence, the ``arbitrary" constant $S_0$ is given as:
\be
S_{0}=S_{BH}\Big[1-\sqrt{1-\frac{J^2}{4 r^{2}_{in}}}
 -\frac{J}{2 r_{in}}\arcsin(\frac{J}{2 r_{in}})\Big]\, .
\label{constant}
\ee
\par\noindent
Therefore, substituting expression (\ref{constant}) 
for the ``arbitrary" constant $S_0$ in equation (\ref{mentropy1})
 we get the full expression for the modified entropy due to the 
specific modelling of the self-gravitation effect of the $(2+1)$ BTZ 
black hole:
\bea
S_{bh}(\omega, M, \Lambda, J)=S_{BH}+
4\pi\Big[ r_{out}\sqrt{1-\frac{J^2}{4r^{2}_{out}}}
-r_{in}\sqrt{1-\frac{J^2}{4 r^{2}_{in}}}\Big]\nn\\
+2\pi J\Big[\arcsin\left(\frac{J}{2 r_{out}}\right)
 -\arcsin(\frac{J}{2 r_{in}})\Big]\, .
\label{mentropy2}
\eea 
\par\noindent
We see that there are deviations from the standard 
result previously mentioned (see equation (\ref{entropy1}))
 for a fixed BTZ black hole background. 
The entropy of the BTZ black hole is no longer the Bekestein-Hawking 
area formula $S_{BH}$.
The corrections to the Bekestein-Hawking area formula for the entropy
 are not obviously negative. In order to examine if the modified entropy 
is smaller or larger compared to the entropy $S_{BH}$ 
for the fixed BTZ black hole background we work to first order in $\omega$.
\par\noindent
We Taylor expand $r_{out}$ and keep up to linear terms in $\omega$. 
The result is:  
\be
r_{out}=r_{in}\left(1-\epsilon\, \omega\right)
\ee
where $\epsilon$ is a parameter:
\be
\epsilon = \frac{1}{2\sqrt{M^2 -\Lambda J^2}}\, .
\ee
This parameter is small for sufficiently 
large mass of the BTZ black hole (as it is expected for such a
 semiclassical analysis here employed since the radiating 
matter is viewed as point particles).
\par\noindent
Therefore, the modified entropy in (\ref{mentropy2}) 
evaluated to first order in $\omega$ is given as:
\be
S_{bh}(\omega, M, \Lambda, J)=S_{BH}\Bigg\{1-\epsilon\,\omega
\Big[1-\frac{J^2}{8 r_{in}^2}\Big]\Bigg\}
\label{mentropy3}
\ee 
where
\be
\frac{J^2}{8 r_{in}^2}\ll 1
\ee
for sufficiently large mass of the BTZ black hole.
\par \noindent
It is now obvious that to first order in $\omega$,
 the semiclassical corrections to the
Bekenstein-Hawking area formula for the entropy are negative. 
Therefore the maximum entropy that a black hole can have is the
Bekenstein-Hawking entropy, $S_{BH}$, and when 
semiclassical effects are included
the entropy decreases:
\be
S_{bh}(\omega, M, \Lambda, J)\leq S_{BH}\, .
\label{bound}
\ee
A welcomed but not unexpected result is that 
equality in the above equation holds for zeroth order in 
$\omega$ .  
\par \noindent
We also see  that the modified entropy (\ref{mentropy3}) 
of the ($2+1$) BTZ black hole is proportional to its horizon area. 
The modified entropy (\ref{mentropy3}) will then be written as:
\bea
S_{bh}(\omega, M, \Lambda, J)&=& \alpha S_{BH}\nn\\
&=&\frac{\alpha}{4}\mathcal{A}_{H}
\eea
where the dimensionless proportionality factor $\alpha$ :
\be
\alpha =\Bigg\{1-\epsilon\,\omega
\Big[1-\frac{J^2}{8 r_{in}^2}\Big]\Bigg\}
\ee
 is a function of the emitted shell's energy $\omega$ and the ($2+1$) 
BTZ black hole parameters: (i) mass, (ii) angular momentum. 
Therefore, the modified entropy can be written as:
\be
S_{bh}(\omega, M, \Lambda, J)=\frac{1}{4}\mathcal{A}_{bh}(\omega, M, \Lambda, J)
\ee
where the area of the ($2+1$) BTZ black hole has now been modified to:
\be
\mathcal{A}_{bh}(\omega, M, \Lambda, J)=\alpha \mathcal{A}_{BH}.
\ee
This is mainly the reason for the modifications to the Bekestein-Hawking entropy 
of the ($2+1$) BTZ black hole due to the specific modelling of the self-gravitation 
effect adopted here.

\section*{Acknowledgements}
The author would like to thank Ass. Professor T. Christodoulakis
for useful discussions and enlightening comments. This work is
financially supported in part by the University of Athens' Special
Account for the Research.



 \end{document}